\documentclass[twocolumn]{jpsj3}
\usepackage{color}

\title{Spectroscopic Study of $^{75}$As and $^{139}$La NMR on Layered Structure Ferromagnet LaCoAsO}

\author{Hiroto OHTA\thanks{E-mail address:shioshio@kuchem.kyoto-u.ac.jp}, Chishiro MICHIOKA, and Kazuyoshi YOSHIMURA\thanks{E-mail address:kyhv@kuchem.kyoto-u.ac.jp} 
}

\inst{
Department of Chemistry, Graduate School of Science, Kyoto University, Kyoto 606-8502, Japan
}

\abst{
$^{75}$As and $^{139}$La field-swept NMR spectra were obtained for the novel weakly itinerant ferromagnet LaCoAsO with a two-dimensional layered structure above the Curie temperature of 55 K.
By analyzing NMR spectra, temperature dependences of Knight shift $K$ and nuclear quadrupole resonance frequency $\nu _Q$ were obtained successfully for each nucleus.
We confirmed from the so-called $K$-$\chi$ plots that the macroscopic magnetization of our \mbox{LaCoAsO} powder sample is intrinsic and does not contain the contribution from impurity phases.
We estimated hyperfine coupling constants from the slope of $K$-$\chi$ plots and compared them with that of an iron-arsenide superconductor.
}

\kword{%
cobalt pnictide, itinerant ferromagnet, two-dimensional,  NMR spectroscopy
}

\begin{document}
\maketitle

\section{Introduction}
Recently, one of the most important keywords for the strongly correlated electron systems has been the ``layered structure".
After the discovery of high-$T_{\textrm{c}}$ copper oxide superconductors, dozens of compounds with a layered structure have been discovered and intensively studied, especially  in cases of superconductors, e.g., Sr$_2$RuO$_4$\cite{Sr2RuO4}, CeCoIn$_5$\cite{CeCoIn5}, and $bi$layer hydrated Na$_x$CoO$_2\cdot y$H$_2$O\cite{Takada_Nature}.
Some novel phenomena including unconventional superconductivity occur because of 2-dimensionality, which is the reason why the layered structure is important.
The recent discovery of iron-pnictide superconductors\cite{Kamihara_LaFeAsOF} reminded us of the importance of this keyword.

\begin{figure}
\begin{center}
\includegraphics[width=6.5cm]{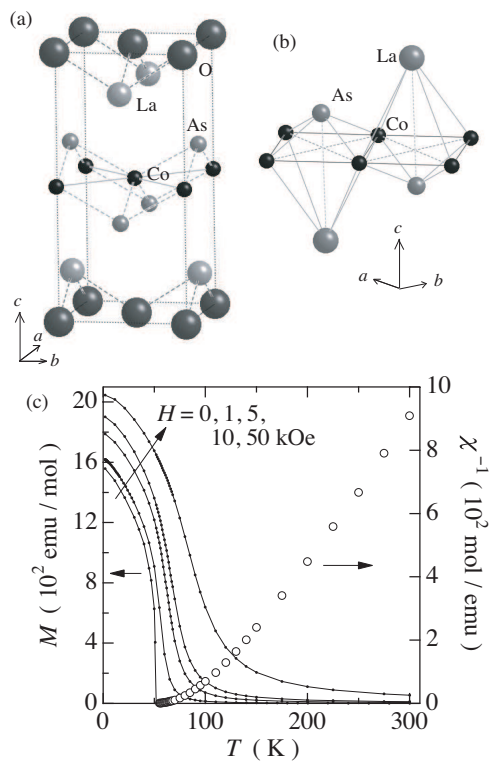}
\end{center}
\caption{
(a) Crystal structure of \mbox{LaCoAsO}.
The space group is $P4/nmm$ (tetragonal).
Lattice parameters $a$ = $b$ = 4.055\hspace{0.5em}\AA\hspace{0.5em} and $c$ = 8.462\hspace{0.5em}\AA.
(b) Frameworks of As-Co and La-Co networks.
Point symmetries of As and La are both $4mm$.
As and La sites are 0.1321 and 0.3491 higher than the Co site along the $c$-axis in the unit of $c$.
(c) Temperature dependences of $M$ at various $H$ and $\chi ^{-1}$ values.
}
\label{CrystalStructure}
\end{figure}

Iron-pnictide superconductors have rich variation in the related compounds.
For now, three systems denoted by 1111 (e.g., LaFeAsO$_{1-x}$F$_x$\cite{Kamihara_LaFeAsOF}), 122 (e.g., Ba$_{1-x}$K$_x$Fe$_2$As$_2$\cite{BaKFe2As2}), and 11 (e.g., Fe$_{1+x}$Se\cite{FeSe}) are intensively studied, and other systems, for example, Li$_{1-x}$FeAs\cite{LiFeAs} and Sr$_4$Sc$_2$Fe$_2$P$_2$O$_6$\cite{Sr42226}, are also known as the members of the family.
They all have the common structure of FeAs (FeSe) conduction planes, where irons form a two-dimensional (2D) square lattice.
From the point of view of itinerant magnetism, we do not need to restrict the target of research to iron compounds.
It has been reported that iron sites of LaFeAsO can be substituted by other transition metals.
Studies on the synthesis and structural analyses of $Ln$$M$$Pn$O ($Ln$: lanthanoids, $M$: transition metals, $Pn$ = P and As) were first reported by certain German groups\cite{Zimmer}, and several reports on physical properties followed them\cite{Nientiedt_LaMnPO, Watanabe_LaNiPO, Yanagi_LaCoPnO, Sefat_LaCoAsO, Ohta_LaCoAsO}.
These compounds may help us to study the 1111  system of iron arsenide superconductors since they have the same structure and their electronic structures sre expected to be related to each other.

In the case of $M$ being cobalt, LaCoAsO and LaCoPO were reported to be itinerant ferromagnets with Curie temperatures ($T_{\textrm{C}}$) of about 50 and 60 K, respectively\cite{Yanagi_LaCoPnO, Sefat_LaCoAsO, Ohta_LaCoAsO}.
In our previous report\cite{Ohta_LaCoAsO}, we analyzed the magnetization of \mbox{LaCoAsO} using the theories of spin fluctuations\cite{Moriya_SCR,Takahashi_SpinFluctuations}, and we pointed out that the dimensionality of spin fluctuations near $T_{\textrm{C}}$ is three and that 2D spin fluctuations above 150 K seem to dominate over 3-dimensional ones.
Such a quasi-2D itinerant ferromagnet has hardly been studied thus far from the experimental point of view.
Recently, a $^{31}$P NMR study on LaCoPO has been reported\cite{Sugawara_LaCoPO}, in which the 2D ferromagnetic nature is claimed to be observed in the temperature region higher than $T_{\textrm{C}}$.
We also reported the magnetovolume effects on CoAs planes of \mbox{LaCoAsO} by substituting La for Ce $\sim$ Gd\cite{Ohta_LnCoAsO}, in which we discussed the highly 2D ferromagnetic nature of \mbox{LaCoAsO} from unconventional behaviors of $T_{\textrm{C}}$ and the spontaneous magnetization of the ground state ($P_{\textrm{s}}$).

In this report, we presented the results of $^{75}$As and $^{139}$La NMR spectral measurements, and analyses of spectra of the itinerant ferromagnet \mbox{LaCoAsO}.
We showed the validity of our analyses in the previous report\cite{Ohta_LaCoAsO} by comparing Knight shifts with macroscopic magnetic susceptibilities.
We also discussed the results of analyses by comparing them with those of iron-pnictide superconductors.

\section{Experimental Methods}
We synthesized polycrystalline samples of LaCoAsO by solid-state reaction methods from powders of La (purity: 99.9\%), As (99.99\%), and CoO (99.99\%).
The detailed synthesis conditions are described in our previous report\cite{Ohta_LaCoAsO}.
By powder X-ray diffraction measurements, we confirmed our polycrystalline sample to be in a single phase of \mbox{LaCoAsO} with the crystal structure shown in Fig. \ref{CrystalStructure}(a).
We present the temperature dependences of magnetizations ($M$) at various magnetic fields ($H$) and inverse magnetic susceptibilities ($\chi ^{-1}$) in Fig. \ref{CrystalStructure}(c)\cite{Ohta_LaCoAsO, Ohta_LnCoAsO}.
These data were obtained using the magnetic property measurement system (MPMS) (Quantum Design Inc.) installed in the Research Center for Low Temperature and Materials Sciences, Kyoto University.
The NMR spectra were obtained by integrating the spin-echo signals while sweeping $H$ with the frequency fixed to $\nu _{exp}$ = 36.46 MHz.
The polycrystalline sample without any magnetic orientations was used for the measurements.
The $M$ of \mbox{LaCoAsO} was measured by changing the magnetic field $H$ up to 14 T using a physical property measurement system (PPMS) (Quantum Design Inc.) in a temperature range between 60 and 300 K.

\begin{figure}
\begin{center}
\includegraphics[width=8.3cm]{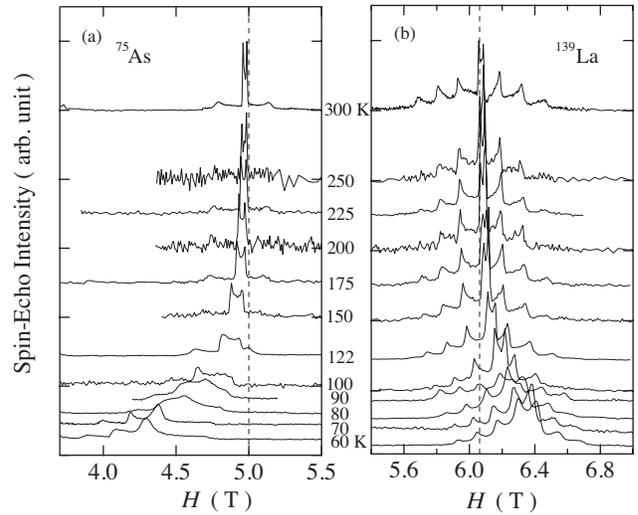}
\end{center}
\caption{
Field-swept NMR spectra of (a)$^{75}$As and (b)$^{139}$La nuclei of \mbox{LaCoAsO} at the fixed frequency $\nu _{exp}$ = 36.46 MHz.
Dashed lines in each panel indicate the resonance field without a shift.
}
\label{spe}
\end{figure}

\begin{figure}
\begin{center}
\includegraphics[width=6.5cm]{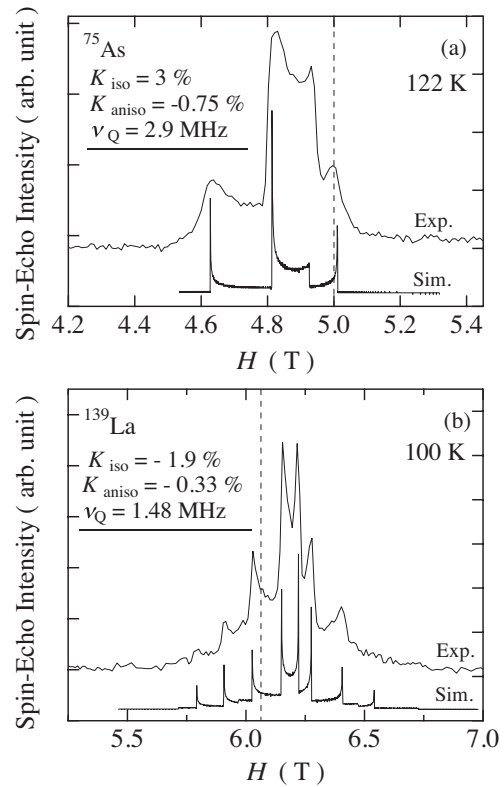}
\end{center}
\caption{
Experimental and simulated spectra of (a) $^{75}$As and (b) $^{139}$La nuclei at 122 and 100 K, respectively.
Dashed lines indicate the zero shift position of each nucleus.
}
\label{spe_sim}
\end{figure}

\section{Experimental Results}
We swept $H$ from 3 to 7 T and successfully observed two distinct spectra at approximately 5 and 6 T, as shown in Fig. \ref{spe}.
Since the gyromagnetic ratios $\gamma$'s of $^{75}$As and $^{139}$La nuclei are 7.292 and 6.0146 MHz/T, and thus, the zero shift fields ($H_0$ ($\equiv \nu _{exp}/\gamma$)) for  $^{75}$As and $^{139}$La nuclei are 5.000 and 6.0619 T, respectively, with $\nu _{exp}$ = 36.46 MHz, we regarded the two spectra at approximately 5 and 6 T as those derived from $^{75}$As and $^{139}$La nuclei, respectively, as observed in Fig. \ref{spe}.
Here, dashed lines in each panel indicate the $H_0$ for each nucleus.
At 300 K, $^{75}$As and $^{139}$La spectra have a structure with a sharp peak corresponding to the central line and two and six satellite peaks, respectively.
Since the nuclear spins $I$ of $^{75}$As and $^{139}$La nuclei are 3/2 and 7/2, respectively, their spectral structures are consistent with that in the case of a large electric field gradient (EFG).
The split of the central line observed for both nuclei is mainly due to the second-order perturbation effect of nuclear quadrupole resonance (NQR).
The observation of the clear spectral structure shows that our sample has a high quality and crystallizes well in each polycrystalline grain.

The spectra of both $^{75}$As and $^{139}$La nuclei show large shifts at low temperatures, and the directions of the shifts are opposite, as observed in Fig. \ref{spe}.
The broadening of the structure with decreasing temperature is also observed in the figure.
Such a broadening can be considered to be due to the effect of the increase in magnetization.
In addition to spectral broadening, there are obvious changes in structure with decreasing temperature, especially in the case of $^{75}$As NMR.
To analyze complicated spectral structures, we simulated spectra including the central line and all the satellite lines of the randomly oriented powders with isotropic and anisotropic Knight shifts ($K_{\textrm{iso}}$ and $K_{\textrm{aniso}}$), and NQR frequency ($\nu _Q$) as parameters for each nucleus case, and compared the results of simulations with the experimental ones.
For the simulation of the spectra, we used the formula in the case of second-order perturbation effects of the nuclear quadrupole interaction being considered and with the assumption that the principal axes of the Knight shift and EFG are parallel to each other\cite{MetalicShift} and to the $c$-axis.
We also assumed for the simulations that the asymmetric parameter within the $ab$-plane ($\eta$) is zero since the symmetry of the crystal structure is tetragonal.

Figures \ref{spe_sim}(a) and \ref{spe_sim}(b) show the typical results of whole views of the simulated and observed spectra of $^{75}$As and $^{139}$La, respectively.
All the singularities in the experimental spectra were well explained by the simulated ones with the parameters shown in each panel.
These results indicate that our sample exhibits a completely random orientation against $H$.
In the case of the $^{139}$La nucleus, $K_{\textrm{iso}}$ and $K_{\textrm{aniso}}$ are both negative, while in the case of the $^{75}$As nucleus, $K_{\textrm{aniso}}$ has a sign opposite to that of $K_{\textrm{iso}}$.

\begin{figure}
\begin{center}
\includegraphics[width=7.4cm]{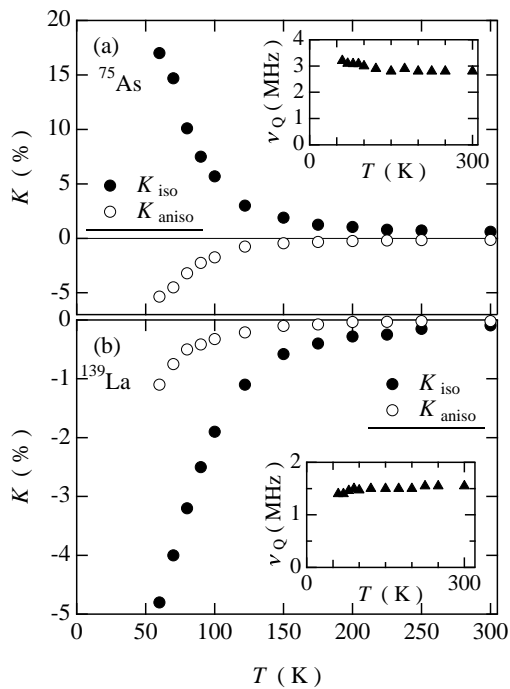}
\end{center}
\caption{
Temperature dependences of $K_{\textrm{iso}}$ and $K_{\textrm{aniso}}$ of (a) $^{75}$As and (b) $^{139}$La, respectively.
The inset in each panel shows the temperature dependence of $\nu _Q$. 
}
\label{K_T}
\end{figure}

Figures \ref{K_T}(a) and \ref{K_T}(b) show the temperature dependences of the estimated Knight shifts for $^{75}$As and $^{139}$La nuclei, respectively.
We also showed the estimated $\nu _Q$ for each nucleus in the inset of each panel.
$\nu _Q$ does not show a marked change but shifts slightly with decreasing temperature, showing that the structural and electronic states of the sample do not change markedly above $T_{\textrm{C}}$.
It should be noted that the $\nu _Q$ values of $^{75}$As and $^{139}$La depend on temperature but in opposite directions.
This fact indicates that the local thermal expansions around As and La sites are induced by different mechanisms.
In both nuclear cases, both the magnitudes of $K_{\textrm{iso}}$ and $K_{\textrm{aniso}}$ showed Curie-Weiss-like temperature dependences, indicating that $K$ is well scaled to the macroscopic $\chi$ in this temperature region.
The same temperature dependence of $K$ with $\chi$ shows that the macroscopic magnetization is intrinsic and originates from electrons that are homogeneously distributed over the sample.

\begin{figure}
\begin{center}
\includegraphics[width=8.2cm]{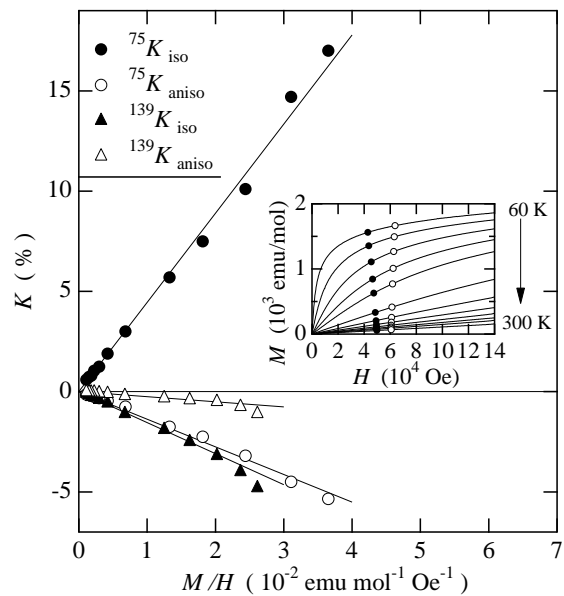}
\end{center}
\caption{
$K$ versus $M/H$ plots with temperature as an implicit parameter.
Solid lines indicate results of linear fitting to the data.
The inset shows $M$ versus $H$ curves at various temperatures from 60 to 300 K.
Closed and open circles indicate the magnetizations at various $H_{res}^{\textrm{iso}}$ values of $^{75}$As and $^{139}$La, respectively. 
}
\label{K_CHI}
\end{figure}

To see how the $K$ and $\chi$ scale with each other, the so-called $K$-$\chi$ plot is also used for nearly or weakly itinerant ferromagnetic compounds\cite{Pd, Yoshimura_YCoAl}.
The $M$ of the ferromagnetic compounds shows a convex behavior against $H$, especially in the temperature range of around $T_{\textrm{C}}$.
For this reason, the magnetic susceptibility $\chi$ (= $M/H|_{H\rightarrow 0}$) is different from $M/H$, and we should use $M/H$ instead of $\chi$ for the $K$-$\chi$ plot.
As observed in the spectra shown in Fig. \ref{spe}, the resonance field changes with temperature in both nuclear cases.
Therefore, we must use the $M$ at the resonance field ($H_{res}$) for the $K$-$\chi$ plot. 
We obtained the $H_{res}$ that corresponds to $K_{\textrm{iso}}$ ($H_{res}^{\textrm{iso}}$) from the relation $K = (H_0 - H_{res})/H_{res}$.
To estimate $M$ at $H_{res}^{\textrm{iso}}$ for each temperature, we measured the $H$ dependence of $M$ up to 14 T.
The obtained results are denoted by the solid lines in the inset of Fig. \ref{K_CHI}.
Solid and open circles show the $M$ values at various $H_{res}^{\textrm{iso}}$ values of $^{75}$As and $^{139}$La, respectively.
Using these estimated $M$ and $H_{res}^{\textrm{iso}}$ values, we obtained the $K$-$\chi$ plots shown in Fig. \ref{K_CHI}.
Closed and open circles show $K_{\textrm{iso}}$ and $K_{\textrm{aniso}}$ of $^{75}$As, and closed and open triangles, $K_{\textrm{iso}}$ and $K_{\textrm{aniso}}$ of $^{139}$La, respectively.
All the Knight shifts show good linearity against $M/H$.
Solid lines indicate the results of linear fitting to the data.
All the linear lines almost passed the origin of the axes.
The orbital part of the Knight shifts is zero or negligibly small compared with their spin part since both As and La only have filled bands and/or closed shells.
We estimated the total diamagnetic susceptibility derived from core electrons of each ion using the calculated values with relativistic corrections as $\chi _{dia}$ = -6.5 $\sim$ -15.3 $\times$ 10$^{-5}$ emu/mol\cite{Diamag, Diamag_2}, which is negligibly small compared with $M/H$. 
Therefore, the orbital contribution to the magnetic susceptibility is also very small in this system.
Here, we would like to stress again the fact that the macroscopic magnetization of our sample can be attributed not to impurity phases but purely to the itinerant electrons of the sample\cite{Ohta_LaCoAsO}, since NMR can selectively detect the intrinsic signals from the sample. 

In general, spin parts of $K$ and $\chi$ are related to each other as $K_s = A_{hf}\chi _s$, where $A_{hf}$ is the hyperfine coupling constant of the spin part at the probing nucleus.
We obtained $A_{hf}$ values of $^{75}$As and $^{139}$La nuclei from the slope of $K$-$\chi$ plots as $^{75}A_{hf}^{iso}$ = 24.8, $^{75}A_{hf}^{aniso}$ = -7.7, $^{139}A_{hf}^{iso}$ = -8.64, and $^{139}A_{hf}^{aniso}$ = -1.41 kOe/$\mu _{\textrm{B}}$.
We can obtain $A_{hf}$ values along the $c$-axis and within the $ab$-plane from isotropic and anisotropic ones using the relations $K_{\textrm{iso}} = (K_c + 2K_{ab})/3$ and $K_{\textrm{aniso}} = (K_c - K_{ab})/3$, where $K_c$ is the Knight shift along the $c$-axis, and $K_{ab}$, the Knight shift within the $ab$-plane.
The $A_{hf}$ values are $^{75}A_{hf}^c$ = 9.4, $^{75}A_{hf}^{ab}$ = 32.5, $^{139}A_{hf}^c$ = -11.4, and $^{139}A_{hf}^{ab}$ = \mbox{-7.23} kOe/$\mu _{\textrm{B}}$.
The anisotropies of the Knight shifts are derived from the anisotropies of the magnetizations and/or hyperfine coupling constants.
Because we used the magnetization of the randomly oriented powder sample, the anisotropies of the Knight shifts are attributed only to the hyperfine coupling constant in our analyses.
To estimate the exact value of the hyperfine coupling constant for each direction, it is necessary to measure the anisotropy of the magnetization using the single-crystal sample and to analyze the data by obtaining $K$-$\chi$ plots for the $c$-axis and the $ab$-plane directions independently.

Although we only know the anisotropies of the Knight shifts, we can describe the hyperfine coupling constant to some extent.
As shown above, the anisotropies of the Knight shifts of $^{75}$As and $^{139}$La are opposite each other.
If the anisotropy of a Knight shift only comes from that of $\chi$ ($M/H$), the anisotropies of the Knight shifts of $^{75}$As and $^{139}$La nuclei must be along the same direction.
Therefore, the anisotropies of hyperfine coupling constants must exist eventually in this system.
The magnetic dipole field from the magnetic moments of cobalt can be the origin of the anisotropies of the hyperfine coupling constants.
The calculated hyperfine coupling constants of the magnetic dipole from the Co site $A_{dip}$ at As and La sites are $^{75}A_{dip}^c$ = -2$(^{75}A_{dip}^{ab})$ = -1.56 kOe/$\mu _{\textrm{B}}$ and $^{139}A_{dip}^c$ = -2$(^{139}A_{dip}^{ab})$ = 1.03 kOe/$\mu _{\textrm{B}}$.
The off-diagonal elements of the $A_{dip}$ matrices become zero owing to the high point symmetry at As and La sites, as observed in Fig. \ref{CrystalStructure}(b).
Anisotropic hyperfine coupling constants of the magnetic dipole $A_{dip}^{aniso}$'s are -0.78 kOe/$\mu _{\textrm{B}}$ for the As site and +0.52 kOe/$\mu _{\textrm{B}}$ for the La site (isotropic parts of the magnetic dipole become zero for both sites).
The value of $A_{dip}^{aniso}$ at the As site is about one order smaller than the value of $^{75}A_{hf}^{aniso}$, while at the La site, although the estimated value of $A_{dip}^{aniso}$ is on the same order as $^{139}A_{hf}^{aniso}$, their signs are opposite each other.
Therefore, the anisotropies of the hyperfine coupling constants at both sites cannot be explained by the dipole fields from the Co site.
The hyperfine coupling constants at the As site are attributed to the hybridization effect between the 4$s$, 4$p$-orbitals of As and 3$d$-orbitals of Co, while at the La site, conduction electrons are considered to mainly contribute to the hyperfine coupling field, since there seems no overlap between the orbitals of La and Co, as observed in  Fig. \ref{CrystalStructure}(a).
We indicated in our previous report that the magnetic ordering is realized by the Rudderman-Kittel-Kasuya-Yoshida (RKKY) interaction between localized magnetic moments at lanthanoid sites and ordered moments at Co sites in the case of $Ln$CoAsO ($Ln$ = Ce $\sim$ Gd)\cite{Ohta_LnCoAsO}.
The NMR results of the $^{139}$La nucleus well support this scenario.
Finally, we note that the anisotropy of the hyperfine coupling field at the pnictide site in the case of LaCoPO (at the P site) is opposite to that in the case of LaCoAsO (at the As site)\cite{Sugawara_LaCoPO}.
Such a difference may be related to differences in physical properties, e.g., the difference in superconducting transition temperature between arsenide and phosphide in the iron-pnictide system.

Next, we would like to compare our results with those of the FeAs-based superconductors. 
In the FeAs-based superconductors and related compounds, the hyperfine coupling constant at the As site has been reported by only a few groups to the best of our knowledge: in the case of the 1111 system, Grafe $et$ $al$. reported it in LaFeAsO$_{0.9}$F$_{0.1}$\cite{Grafe_Ahf}, and in the case of the 122 system, Kitagawa and coworkers reported it in $A$Fe$_2$As$_2$ ($A$ = Ba and Sr) using single-crystal samples\cite{Kitagawa_BaFe2As2, Kitagawa_SrFe2As2}.
The crystal structures of \mbox{LaCoAsO} and LaFeAsO$_{1-x}$F$_x$ are the same, and therefore, the hyperfine coupling constants of these compounds are expected to be similar.
Indeed, the $^{75}A_{hf}^{ab}$ values of LaFeAsO$_{0.9}$F$_{0.1}$ has been reported to be about 25 kOe/$\mu _{\textrm{B}}$\cite{Grafe_Ahf}, which is close to our estimated value of 32.5 kOe/$\mu _{\textrm{B}}$.
However, the anisotropies of the Knight shift is opposite to that of LaCoAsO, i.e., $^{75}K_a$ $<$ $^{75}K_c$ in LaFeAsO$_{0.9}$F$_{0.1}$.
It seems difficult to explain such a difference in the anisotropy of $K$ from the difference between Fe and Co.
We tried to fit the calculated $^{75}$As spectra to the experimental ones of LaFeAsO$_{0.9}$F$_{0.1}$ in Ref. [24].
As the results, we found that the calculated spectra using both positive and negative $K_{aniso}$ values (corresponding to conditions of $K_a < K_c$ and $K_a > K_c$, respectively) can be well fitted to the experimental spectra within the experimental error.
This is because the contribution to the shape of spectra from the $\nu _Q$ is much larger than that from the anisotropy of the hyperfine field.
Therefore, there still exists the possibility that the anisotropy of the hyperfine coupling constant is the same between the cases of LaFeAs(O, F) and LaCoAsO.
On the other hand, the results obtained for $A$Fe$_2$As$_2$ are consistent with our results from both the magnitude and anisotropic viewpoints, i.e., the value of $^{75}A_{hf}^{ab}$ being 26.4 kOe/$\mu _{\textrm{B}}$ for $A$ = Ba (29.3 kOe/$\mu _{\textrm{B}}$ for $A$ = Sr) is larger than that of $^{75}A_{hf}^{c}$ = 18.8 kOe/$\mu _{\textrm{B}}$ for $A$ = Ba (20.4 kOe/$\mu _{\textrm{B}}$ for $A$ = Sr).
The difference in magnitude from our result may in part originate from the fact that the $^{75}A_{hf}$ values for both directions are estimated using $\chi _{ab}$ and $\chi _c$, not $\chi$ of the powder sample.
Since the single crystal is hard to obtain (and the obtained crystal is small compared with that of the 122 system\cite{Ishikado_1111single}) and both microscopic and macroscopic magnetic susceptibilities are small, the precise anisotropy of the hyperfine coupling constant in the FeAs-based 1111 system seems to be hard to obtain as argued above.
For this reason, our results must aid in the study of the FeAs-based 1111 system, though they contain ambiguities in the magnitude of the hyperfine coupling constant.

Finally, we would like to discuss the spin fluctuations.
To observe the ferromagnetic spin fluctuations of the sample, we must study the temperature dependence of nuclear spin-lattice relaxation rate (1/$T_1$), as reported by Sugawara $et$ $al$. in the case of $^{31}$P NMR in LaCoPO\cite{Sugawara_LaCoPO}.
However, we did not succeed in measuring 1/$T_1$ since it is hard to selectively measure the central or satellite transition lines owing to the complex structure of the spectra in the present case.
Thus, we attempted to obtain magnetically oriented samples for this purpose.
However, we failed, possibly indicating that the anisotropy of magnetization is not very strong in LaCoAsO.

\section{Conclusions}
We measured $^{75}$As and $^{139}$La field-swept NMR spectra of the layered structure weakly itinerant ferromagnet \mbox{LaCoAsO} from 60 to 300 K.
We analyzed the spectra and successfully determined temperature dependences of Knight shifts and NQR frequency for each nucleus.
We succeeded in determining the anisotropies of the Knight shifts for both $^{75}$As and $^{139}$La nuclei.
We precisely obtained the $K$-$\chi$ plots for both nuclei and confirmed from the results that the macroscopic magnetization of \mbox{LaCoAsO} we presented in previous report\cite{Ohta_LaCoAsO} is intrinsic, eliminating the possibility of the contribution of impurity phases to the magnetization.
The hyperfine coupling constants estimated from the slope of the $K$-$\chi$ plot at the As site are found to be similar to those of $A$Fe$_2$As$_2$ ($A$ = Sr and Ba) in both magnitude and anisotropy, indicating that the hyperfine coupling constant in the $ab$-plane is larger than that along the $c$-axis in transition metal arsenides.

\section*{Acknowledgements}
We would like to thank Dr. Y. Nakai and Prof. K. Ishida for fruitful discussion.
This work is supported by Grants-in-Aid for Scientific Research on Priority Area ``Invention of anomalous quantum materials", from the Ministry of Education, Culture, Sports, Science and Technology of Japan (16076210), and also by Grants-in-Aid for Scientific Research from the Japan Society for Promotion of Science (19350030).


\begin{thebibliography}{99}

\bibitem{Sr2RuO4} Y. Maeno, H. Hashimoto, K. Yoshida, S. Nishizaki, T. Fujita, J.G. Bednorz, and F. Lichtenberg: Nature (London) \textbf{372} (1994) 532.

\bibitem{CeCoIn5} H. Hegger, C. Petrovic, E. G. Moshopoulou, M. F. Hundley, J. L. Sarrao, Z. Fisk, and J. D. Thompson: Phys. Rev. Lett. \textbf{84} (2000) 4986.

\bibitem{Takada_Nature} K. Takada, H. Sakurai, E. Takayama-Muromachi, F. Izumi, R. A. Dilanian and T. Sasaki: Nature \textbf{422} (2003) 53.

\bibitem{Kamihara_LaFeAsOF} Y. Kamihara, T. Watanabe, M. Hirano, and H. Hosono: J. Am. Chem. Soc. \textbf{130} (2008) 3296.

\bibitem{BaKFe2As2} M. Rotter, M. Tegel, and D. Johrendt: Phys. Rev. Lett. \textbf{101} (2008) 107006.

\bibitem{FeSe} F. C. Hsu, T. Y. Luo, K. W. Yeh, T. K. Chen, T. W. Huang, P. M. Wu, Y. C. Lee, Y. L. Huang, Y. Y. Chu, D. C. Yan, and M. K. Wu: Proc. Natl. Acad. Sci. USA \textbf{105} (2008) 14262.

\bibitem{LiFeAs} X. C. Wang, Q. Q. Liu, Y. X. Lv, W. B. Gao, L. X. Yang, R. C. Yu, F. Y. Li, and C. Q. Jin: Solid State Commun. \textbf{148} (2008) 538.

\bibitem{Sr42226} H. Ogino, Y. Matsumura, Y. Katsura, K. Ushiyama, S. Horii, K. Kishio, and J. Shimoyama: Supercond. Sci. Technol. \textbf{22} (2009) 075008.

\bibitem{Zimmer} For example, B. I. Zimmer, W. Jeitschko, J. H. Albering, R. Glaum, and M. Reehuis: J. Alloys Compd. \textbf{229} (1995) 238, and the references therein.

\bibitem{Nientiedt_LaMnPO} A. T. Nientiedt, W. Jeitschko, P. G. Pollmeier, and M. Brylak: Z. Naturforsch. \textbf{52b} (1997) 560.

\bibitem{Watanabe_LaNiPO} T. Watanabe, H. Yanagi, T. Kamiya, Y. Kamihara, H. Hiramatsu, M. Hirano, and H. Hosono: Inorg. Chem. \textbf{46} (2007) 7719.

\bibitem{Yanagi_LaCoPnO} H. Yanagi, R. Kawamura, T. Kamiya, Y. Kamihara, M. Hirano, T. Nakamura, H. Osawa, and H. Hosono: Phys. Rev. B \textbf{77} (2008) 224431.

\bibitem{Sefat_LaCoAsO} A. S. Sefat, A. Huq, M. A. McGuire, R. Jin, B. C. Sales, D. Mandrus, L. M. D. Cranswick, P. W. Stephens, and K. H. Stone: Phys. Rev. B \textbf{78} (2008) 104505. 

\bibitem{Ohta_LaCoAsO} H. Ohta and K. Yoshimura: Phys. Rev. B \textbf{79} (2009) 184407.

\bibitem{Moriya_SCR} T. Moriya: $Spin$ $Fluctuations$ $in$ $Itinerant$ $Electron$ $Magnetism$ (Springer-Verlag, New York, 1985).

\bibitem{Takahashi_SpinFluctuations} Y. Takahashi: J. Phys. Soc. Jpn. \textbf{55} (1986) 3553.

\bibitem{Sugawara_LaCoPO} H. Sugawara, K. Ishida, Y. Nakai, H. Yanagi, T. Kamiya, Y. Kamihara, M. Hirano, and H. Hosono: J. Phys. Soc. Jpn. \textbf{78} (2009) 113705.

\bibitem{Ohta_LnCoAsO} H. Ohta and K. Yoshimura: Phys. Rev. B \textbf{80} (2009) 184409.

\bibitem{MetalicShift} G. C. Carter, L. H. Bennett, and D. J. Kahan: $Metalic$ $Shifts$ $in$ $NMR$, Part I (Pergamon Press, Oxford, 1977) 79.

\bibitem{Pd} J. A. Seitchik, A. C. Gossard, and V. Jaccarino: Phys. Rev. \textbf{136} (1964) A1119.

\bibitem{Yoshimura_YCoAl} K. Yoshimura, M. Takigawa, Y. Takahashi, H. Yasuoka, and Y. Nakamura: J. Phys. Soc. Jpn. \textbf{56} (1987) 1138.

\bibitem{Diamag} $Landolt$ - $B\ddot{\textrm{a}}rnstein$, Neue Serie I\hspace{-.1em}I/11 (Springer-Verlag, New York, 1981) 27.

\bibitem{Diamag_2} L. B. Mendelsohn, F. Biggs, and J. B. Mann: Phys. Rev. A \textbf{2} (1970) 1130.

\bibitem{Grafe_Ahf} H. -J. Grafe, D. Paar, G. Lang, N. J. Curro, G. Behr, J. Werner, J. Hamann-Borrero, C. Hess, N. Leps, R. Klingeler, and B. Buchner: Phys. Rev. Lett. \textbf{101} (2008) 047003.

\bibitem{Kitagawa_BaFe2As2} K. Kitagawa, N. Katayama, K. Ohgushi, M. Yoshida, and M. Takigawa: J. Phys. Soc. Jpn. \textbf{77} (2008) 114709.

\bibitem{Kitagawa_SrFe2As2} K. Kitagawa, N. Katayama, K. Ohgushi, and M. Takigawa: J. Phys. Soc. Jpn. \textbf{78} (2009) 063706.

\bibitem{Ishikado_1111single} M. Ishikado, S. Shamoto, H. Kito, A. Iyo, H. Eisaki, T. Ito, and Y. Tomioka: Physica C \textbf{469} (2009) 901.

\end{thebibliography}
\end{document}